# Controlling spatiotemporal nonlinearities in multimode fibers with deep neural networks

**Uğur Teğin**[1,2,*], **Babak Rahmani**[1], **Eirini Kakkava**[2], **Navid Borhani**[2], **Christophe Moser**[1], **and Demetri Psaltis**[2]

[1]Laboratory of Applied Photonics Devices, École Polytechnique Fédérale de Lausanne, Lausanne, Switzerland
[2]Optics Laboratory, École Polytechnique Fédérale de Lausanne, Lausanne, Switzerland
[*]ugur.tegin@epfl.ch

## ABSTRACT

Spatiotemporal nonlinear interactions in multimode fibers are of interest for beam shaping and frequency conversion by exploiting the nonlinear propagation of different pump regimes from quasi-continuous wave to ultrashort pulses centered around visible to infrared pump wavelengths. The nonlinear effects in multi-mode fibers depend strongly on the excitation condition, however relatively little work has been reported on this subject. Here, we present the first machine learning approach to learn and control the nonlinear frequency conversion inside multimode fibers by tailoring the excitation condition via deep neural networks. Trained with experimental data, deep neural networks are adapted to learn the relation between the spatial beam profile of the pump pulse and the spectrum generation. For different user-defined target spectra, network-suggested beam shapes are applied and control over the cascaded Raman scattering and supercontinuum generation processes are achieved. Our results present a novel method to tune the spectra of a broadband source.

## Introduction

Multimode fibers (MMFs) have found applications in several fields in the last decades mainly in telecommunication and imaging[1,2]. In recent years, spatiotemporal nonlinearities in MMFs, have become the subject of strong interest in various fundamental and applied areas, from single-pass propagation to spatiotemporally mode-locked laser cavities[3-5]. In the single-pass nonlinear propagation studies, by using different pulse types from visible to IR wavelengths with a Gaussian beam profile numerous interesting phenomenon[6-15] including spatiotemporal instability, self-beam cleaning and supercontinuum generation are reported with graded-index multimode fibers (GRIN MMFs). Although, for most of these spatiotemporal nonlinear phenomena the importance of the excitation condition is emphasized, much remains to be done before we can fully understand the impact of the excitation condition on the nonlinear spatiotemporal propagation and to control it subsequently.

In recent years, spatial control of light propagation in MMFs has been widely achieved, *in the linear regime*, via iterative methods, phase conjugation and the transmission matrix method for applications such as imaging and material processing[16,17]. Recently, machine learning tools have been proposed to simplify the calibration and improve the robustness of the system in the absence of optical nonlinearities[18]. Deep neural networks (DNNs) proved useful for classification/ reconstruction of the information sent to km-long MMFs solely from the intensity measurement at the output of the fibers[19,20]. For the nonlinear propagation regime, only adaptive algorithms have been brought forth and been shown to be successful in harnessing the entangled spatiotemporal nonlinearities such as Kerr beam self-cleaning of low-order modes[21] and optimization of the intensity of targeted spectral peaks generated by Raman scattering or four-wave

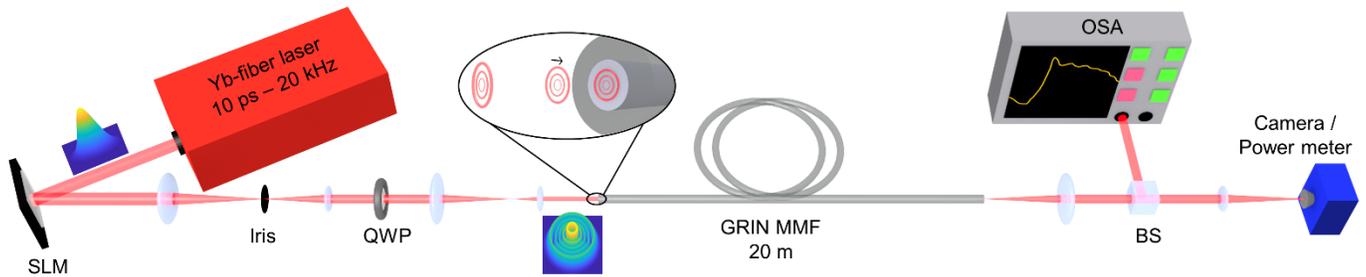

**Figure 1. Experimental setup.** SLM spatial light modulator, QWP quarter-wave plate, BS beam splitter, OSA optical spectrum analyzer.

mixing[22].

In this article, we report the results of our studies on the effect of initial excitation condition of the GRIN MMFs to nonlinear spatiotemporal propagation by employing machine learning approaches. Specifically, we achieved control over multimodal nonlinear frequency conversion dynamics and demonstrated that spatiotemporal nonlinear pulse propagation can be learned by DNNs. Once trained, the DNNs can predict the necessary beam shape for the pump pulses to produce a desired spectral shape within the limitations of the triggered nonlinear effects at the end of the GRIN MMF. In particular, we showed that cascaded stimulated Raman scattering (SRS) based broadening of the spectrum as well as supercontinuum generation, two highly nonlinear phenomena, can be experimentally controlled for the first time in the literature with machine learning tools.

## Results

### Experimental setup and dataset collection

In our experiments (Fig. 1), we launched 10 ps pulses centered around 1030 nm with adjustable peak power into a 20 m GRIN MMF with 62.5 µm core diameter. A phase-only spatial light modulator (SLM), 8f imaging system and a quarter-wave plate is placed before the GRIN MMF. Here, we study two particular phenomena by adjusting the peak power of the pump pulses to either 85 kW or 150 kW. In the first case, spectral broadening by cascaded SRS while in the second case supercontinuum generation are recorded at the output of the GRIN MMF by using a pump pulse with Gaussian beam profile. Cascaded SRS based spectral broadening in GRIN MMFs has been extensively studied in the past and spatiotemporal pulse propagation is presented as the leading mechanism for the cascaded Raman Stokes generation. In GRIN MMFs, Raman Stokes peaks are reported with different beam shapes which lead to different propagation paths for each Raman peak. By compensating the chromatic dispersion difference, the multimode propagation enhances the cross-phase modulation between the Raman peaks, hence triggering the generation of a supercontinuum formation for higher peak powers after reaching to the zero-dispersion wavelength (ZDW)[13,15].

Numerical calculations are performed to determine the suitable approach to define the preliminary excitation patterns for the experimental studies (see Supplementary Note 1 for details) which revealed significant spectral differences entirely due to the initial power distribution between the fiber modes. Similar behavior was numerically reported in the literature for multimode holey fibers[23]. The propagation differences observed in numerical studies (Supplementary Fig. 1) can be understood by studying the nonlinear coupling term used in the numerical calculations. According to the nonlinear coupling between the modes determined by the overlap integrals, some of the intermodal processes are favored by different modal symmetry classes and particular modes act as a pump for these nonlinear effects. In our calculations, we found that by initially favoring high order modes in spatiotemporal nonlinear propagation, a medium to generate broad output spectrum can be achieved since excitation of the higher-order modes creates an environment which encourages the nonlinear intermodal interactions. In the literature, the importance of the beam size on the fiber facet to initiate multimode propagation is emphasized in experimental studies related with self-beam cleaning and spatiotemporal instability[6,12].

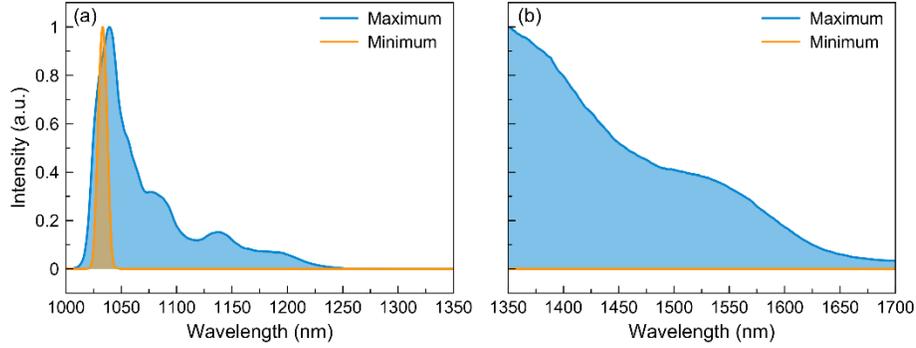

**Figure 2. Difference in output spectra measured in the datasets.** Maximum and minimum frequency conversion measured for 3000 different excitation conditions for 85 kW peak power (a) and for 150 kW peak power (b).

To experimentally study the effect of excitation condition to nonlinear pulse propagation, a set of beam profiles containing 3000 samples are calculated by superposing pre-defined base patterns with random non-repeating amplitudes from 0 to 1 and fixed sum for each candidate beam shape. Guided by simulations, a mixture of the 5 lowest order $LG_{4X}$ modes are selected as the base patterns to create beam profiles of the pump pulses. To minimize the power level changes due to beam shaping, the number of coefficients is intentionally limited as 5 and the $LG_{40}$ mode is added to each calculated beam shape as a background. Here we would like to emphasize that the selected LG modes are calculated for the free-space propagation and the energy distribution of these patterns cannot be directly related to the energy distribution of the modes supported by the fibers. The complex amplitude modulation method is applied to calculate the required phase patterns to generate the designed beam profiles with the phase-only SLM[24]. The achieved beam profiles are ~ 80X demagnified with the imaging system to excite GRIN MMF.

Pulses with 85 kW and 150 kW peak powers and different beam shapes impinge on the GRIN MMF facet while the output spectra are being recorded for each applied pattern. For both cases of interest, strong variations at the output spectra are observed. For 150 kW pump peak power, as the spectrum approaches ZDW, the driving nonlinear effect changes and instead of cascaded SRS, modulation instability based spectral broadening occurs. Therefore, we focused on wavelength range above the ZDW (1350 nm to 1700 nm)[13,15]. The most extreme cases recorded in the datasets are presented in Fig. 2.

**Machine learning and controlling nonlinearities**

For both peak power levels, the machine learning approach is employed to analyze the experimentally collected datasets. In both cases, the same network architecture employed is comprised of four hidden layers as demonstrated in Fig. 3. Measured spectra are fed to the network as inputs and for each spectrum, the coefficients to generate the corresponding beam profiles are labeled as the output variables. DNNs can learn the spatiotemporal nonlinear pulse propagation inside the tested GRIN MMF by adjusting their weights to learn the relation between the generated spectra and the excitation condition as described by the coefficients used to determine the input beam shape. The training and validation results of the DNNs are demonstrated in Supplementary Note 2.

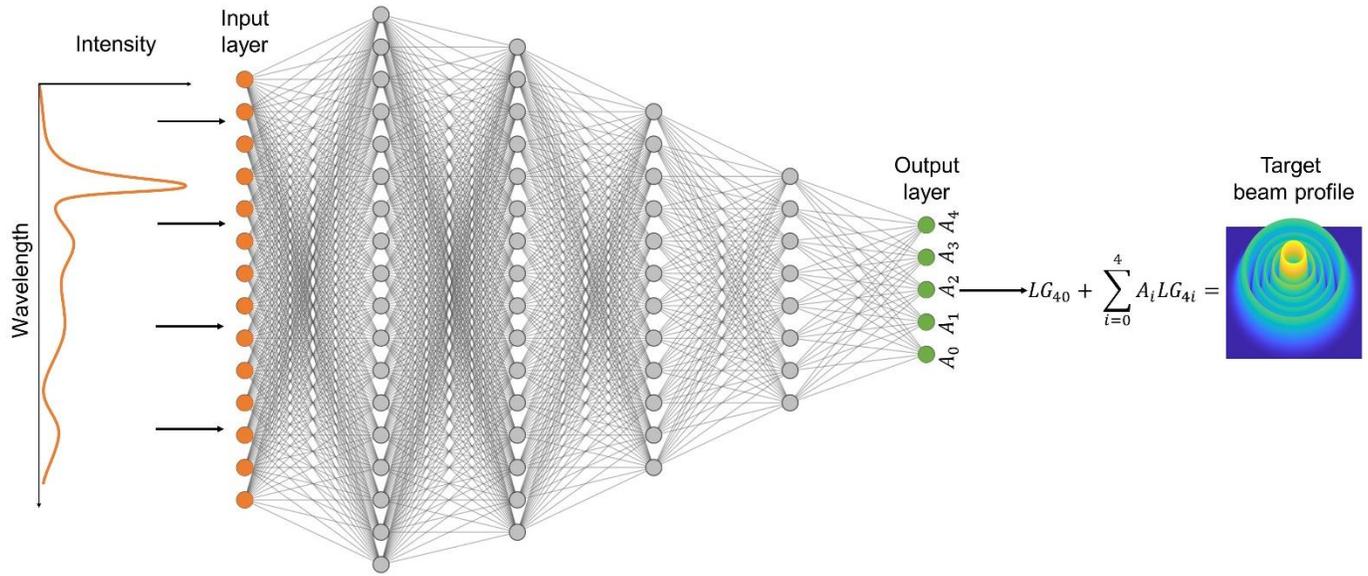

**Figure 3. Schematic of the learning spatiotemporal nonlinear pulse propagation.** Each input spectrum is linked via the proposed neural network to its corresponding coefficients at the output of the network which in turn is used to generate the required beam profile corresponding to the input spectrum.

To experimentally investigate the performance of the trained DNN, the control on the output spectra is demonstrated. For this, a collection of synthetic spectral shapes is generated via summations of Gaussian distributions with different amplitudes and widths. It should be noted that the synthetic spectra should still lie within the limits of the recorded spectral dataset. These target spectra are then feed to the DNN to predict the required beam shapes of the target pump pulses. For each designed target spectra, the DNN predicts the coefficients of the $LG_{4x}$ patterns and from these coefficients, the required input beam shapes. Here, we rely on the ability of the neural network to generalize since the user-defined spectra is not from the test nor the training data set. For 85 kW pump peak power level, summations of different Gaussian functions centered around the Raman Stokes peaks of the silica medium are used to design the targeted spectra. Results of the tests for controlling cascaded SRS generation are presented in Fig. 4. The measured spectra corresponding to the DNN-predicted beam shapes are well matched with the target spectra (Fig. 4a-c). For 150 kW pump peak power level, spectral shapes are designed based on the trailing edge of a Gaussian function centered around 1350 nm since the targeted wavelength range corresponds to the trailing edge of the supercontinuum spectra. The same procedure explained for 85 kW pump peak power level is applied to control the supercontinuum generation. Targeted and experimentally measured spectra are presented in Fig. 5a-c and the experimentally measured spectra which are in good agreement with the targeted spectra. DNN predicted beam profiles of the pump pulses are shown in insets of Fig. 5d-f.

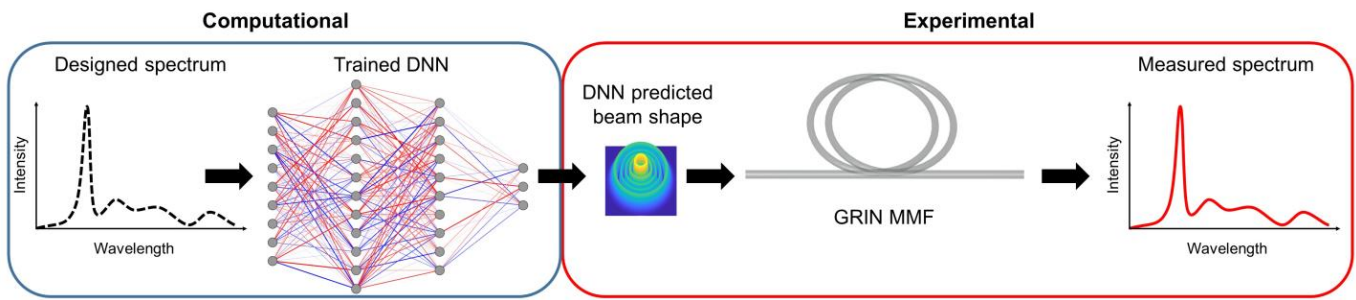

**Figure 4. Schematic of the control experiments.** To experimentally generate the computationally designed spectra, trained DNNs are used to estimate the required beam shape of the pump pulses.

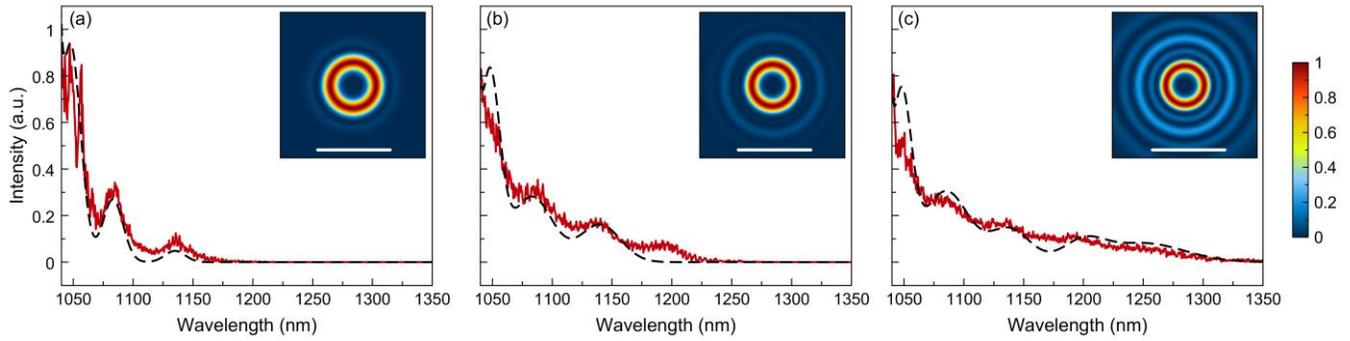

**Figure 5. Controlling cascaded SRS generation with DNN.** a-c Designed (dashed) and measured output spectra for 85 kW peak powers. The insets show DNN suggested beam shapes for pump pulses to generate designed spectra. Scale bars indicated for DNN suggested beam shapes are 3 mm.

## Discussion

The main result of this article is that machine learning tools can reveal the underlying basis of the spatiotemporal nonlinear propagation in MMFs which have been considered chaotic and complex over the years. Here DNNs are employed to learn the relation between the initial modal energy distribution of the fiber and nonlinear frequency conversion. Trained with experimental data, our DNNs are shown to be a powerful toolkit to harness the nonlinear dynamics of the MMF within the nonlinear dynamics defined by the system.

Due to non-surjective relation between the initial modal energy distribution of the fiber and nonlinear frequency conversion dynamics, the experimental efficiency of the DNNs is a significant topic. For the both peak power levels (85 kW and 150 kW), more than 50% efficiency is achieved in the control experiments. By introducing artificial noise to designed spectra, the experimental estimation efficiencies of DNNs are improved and reached to 80%. We believe this improvement is due to the noisy nature of the experimentally collected dataset but it needs to be further investigated in future work. We observed that DNNs learned the behavior of initially favoring high order modes to achieve broader output spectrum which is also presented in our numerical simulations. For both of the peak power levels, to generate targeted spectra which requires more nonlinear conversions, DNNs are increased the coefficients of the high order base patterns (Fig 4-inset and Fig 5-inset).

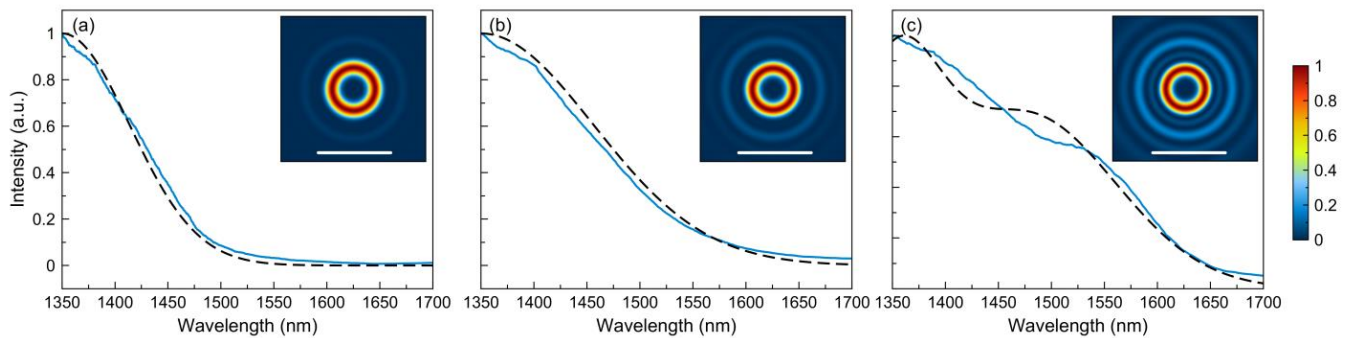

**Figure 6. Controlling supercontinuum generation with DNN.** a-c Designed (dashed) and measured output spectra for 150 kW peak powers. The insets show DNN suggested beam shapes for pump pulses to generate designed spectra. Scale bars indicated for DNN suggested beam shapes are 3 mm.

In conclusion, we showed that spatiotemporal nonlinear pulse propagation can be learned and controlled by machine learning tools. We demonstrated that spectral broadening based on the cascaded SRS and supercontinuum generation can be both altered by tuning the initial modal energy distribution of the fiber through shaping the beam profile of the pump pulse by implementing experimentally trained DNNs. The machine learning approach reported here can be employed to understand and tune other nonlinear effects and relations. Our results present a novel path toward automated tunable fiber-based broadband sources.

## Methods

### Simulations

Simulations are performed with numerically solving multimode nonlinear Schrödinger equation[26]. For numerical integration with high accuracy, we prefer the fourth-order Runge-Kutta in the Interaction Picture method[27]. We used an integration step of 5 µm and time resolution of 2.4 fs with 20 ps time window width. To reduce the computational time, we considered the first 5 linearly polarized modes of the GRIN MMF with 62.5 µm core diameter. Pump pulses centered at 1030 nm with 2 ps duration and 500 kW peak power are numerically propagated for 1 m distance. In our simulations, we included Raman process and third-order dispersion. Relative index difference is assumed as 0.01.

### Experiments

A standard, commercially available GRIN MMF of 20 m length, 62.5 µm core diameter and 0.275 NA (Thorlabs – GIF625) is used in our experiments. Fiber is coiled with 25 cm diameter and rest on the optical table without additional cooling. A laser source (Amplitude Laser - Satsuma) generating pulses at 1030 nm with 10 ps duration and 20 kHz repetition rate is used. Beam shaping to pump pulses are applied with a phase-only SLM (Holoeye Pluto-NIR II). At the end of GRIN MMF, optical spectra are measured with an optical spectrum analyzer (Ando Electric - AQ6317B). The output of the fiber is imaged by a 4f system with ~ 19X magnification and output beam profiles are captured with InGaAs camera (Xenics - Xeva).

### Machine learning

The DNNs are implemented using the Tensorflow 1.14 Python library integrated with Keras 2.2.5 API on Google Colab cloud service which provides an Intel Xeon CPU and Nvidia Tesla K80 GPU for machine learning studies. DNNs are consists of an input layer with 881 nodes, four hidden layers with 2048, 1024, 256 and 32 nodes, respectively; and an output layer with 5 nodes. The number of the nodes at the input and output layers are defined according to the size of the experimentally measured spectra and the number of coefficients used to determine the beam shapes. Rectified linear unit (Relu) and sigmoid activation functions are chosen for the hidden layers and output layer, respectively. To prevent overfitting during the learning process, a dropout method with a 0.4 dropping ratio is applied to the hidden layers.

## Acknowledgements (not compulsory)

The authors thank D. Loterie and P. Hadikhani for fruitful discussions regarding the use of beam shaping and machine learning tools.

## Author contributions statement

U.T. performed the experiments with assistance of E.K. and numerical simulations. Machine learning studies are performed by U.T. and B.R. with assistance of N.B. All authors contributed to interpreting the results obtained and overall project supervision by C.M. and D.P.

# Supplementary Information:
# Controlling spatiotemporal nonlinearities in multimode fibers with deep neural networks

**Uğur Teğin**[1,2,*], **Babak Rahmani**[1], **Eirini Kakkava**[2], **Navid Borhani**[2], **Christophe Moser**[1], **and Demetri Psaltis**[2]

[1]Laboratory of Applied Photonics Devices, École Polytechnique Fédérale de Lausanne, Lausanne, Switzerland
[2]Optics Laboratory, École Polytechnique Fédérale de Lausanne, Lausanne, Switzerland
*ugur.tegin@epfl.ch

## Supplementary Note 1: Numerical results

We numerically investigate the effect of excitation to nonlinear pulse propagation by changing the initial energy splitting ratio between the simulated modes ($LP_{01}$, $LP_{02}$, $LP_{03}$, $LP_{04}$ and $LP_{05}$). To initiate the nonlinear effects in 1 m GRIN MMF, pump pulses in the numerical studies are defined with 2 ps duration and 500 kW peak power. When compared with the experiments, rescaling of the propagation length, peak power and simulated modes is necessary to reduce computational time. Multimode nonlinear Schrödinger equation with Raman scattering term (Eq. 1) is numerically solved.

$$\frac{\partial A_p}{\partial z}(z,t) = i\delta\beta_0^{(p)} A_p - i\delta\beta_1^{(p)} \frac{\partial A_p}{\partial t} - i\frac{\beta_2}{2}\frac{\partial^2 A_p}{\partial t^2} + \frac{\beta_3}{6}\frac{\partial^3 A_p}{\partial t^3}$$
$$+ i\gamma \sum_{l,m,n} \eta_{plmn} \left[ (1-f_R) A_l A_m A_n^* + f_R A_l \int h_R A_m(z, t-\tau) A_n^*(z, t-\tau)\, d\tau \right] \quad (1)$$

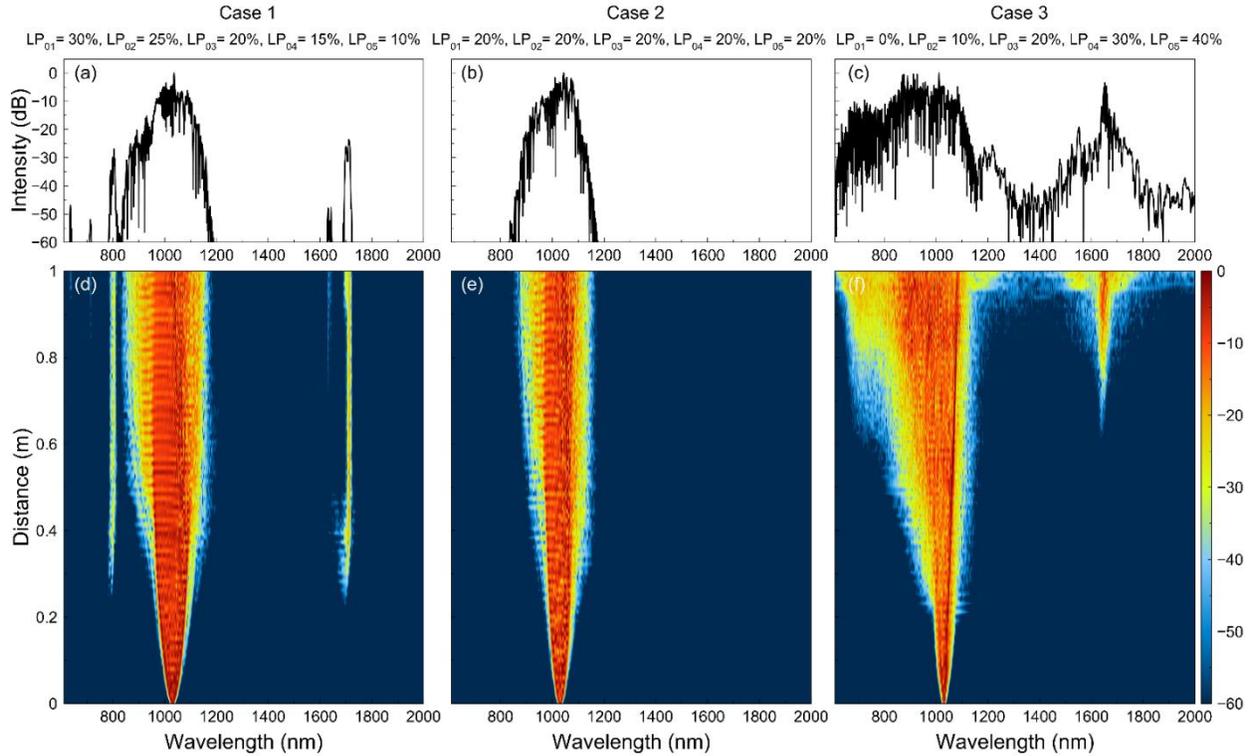

**Supplementary Figure 1.** Simulation results. (a-c) Numerically obtained spectra with different excitation conditions. (d-f) Spectral evolutions through 1 m GRIN MMF.

Here $\eta_{plmn}$ is the nonlinear coupling coefficient, $f_R$ is the fractional contribution of the Raman effect, $h_R$ is the delayed Raman response function and $\delta\beta_0^{(p)}$ ($\delta\beta_1^{(p)}$) is the difference between first (second) Taylor expansion coefficient of the propagation constant for corresponding and the fundamental mode. Fig. 1 demonstrates the variations in the nonlinear pulse propagation with the different initial excitation condition. Even though lower order modes propagate in smaller areas, excitation of the higher-order modes creates an environment which encourages the spatiotemporal nonlinear interactions. As it is shown in Supplementary Fig. 1a, a decent amount of spectral broadening and frequency conversion can be achieved with favoring the fundamental mode (LP01) at energy splitting. For the same pump pulse parameters, the equal excitation of all the modes interestingly resulted in narrower spectra with a less spectral broadening (see Supplementary Fig. 1b). When most of the energy coupled to higher-order modes, highly nonlinear pulse propagation is observed and the optical spectrum evolved to a supercontinuum formation as presented in Supplementary Fig. 1c.

## Supplementary Note 2: Artificial neural network training results

The experimentally collected datasets are divided with a ratio of 0.1 for training (2700 samples) and validation (300 samples). An Adam optimizer is employed to minimize a mean square error (MSE) loss function with the learning rate of $10^{-3}$. Batch type learning with a batch size of 80 is used during the training the DNNs. The change of the MSE (loss) as a function of epoch is presented in Supplementary Fig. 2 for training and validation of 85 kW and 150 kW pump peak power levels.

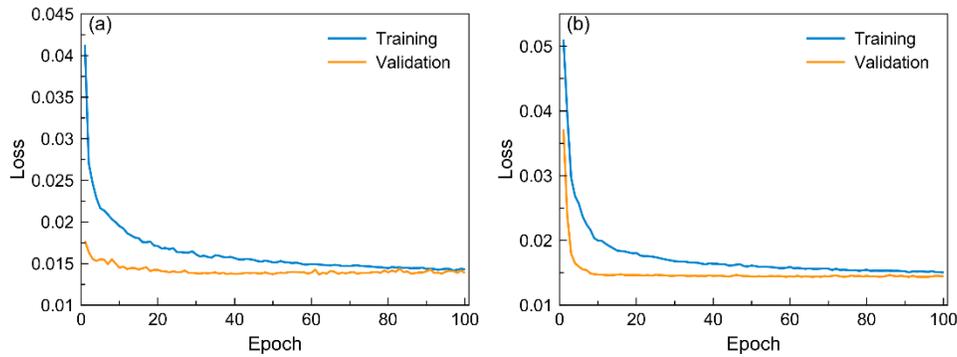

**Supplementary Figure 2.** DNN training results. Change of the loss (MSE error) during training and validation as a function of epoch for cascaded SRS (a) and supercontinuum (b) generation spectra.

## Supplementary Note 3: Experimental Results

Here we present the output beam profiles measured at the end of 20 m GRIN MMF during the controlling spatiotemporal nonlinearities experiments. For 85 kW pump peak power level, the output beam profiles are presented in Supplementary Fig. 3 a-b. The corresponding spectra of the beam profiles are presented in Fig. 5 and with increasing cascaded SRS effect, Raman beam cleanup effect is observed.

For 150 kW pump peak power level, the output beam profiles are presented in Supplementary Fig. 4 a-b. The corresponding spectra of the beam profiles are presented in Fig. 6. When compared with the 85 kW pump peak power level, Raman beam cleanup effect is preserved for the output beam profiles but they contain multimodal behavior featuring larger spatial distribution.

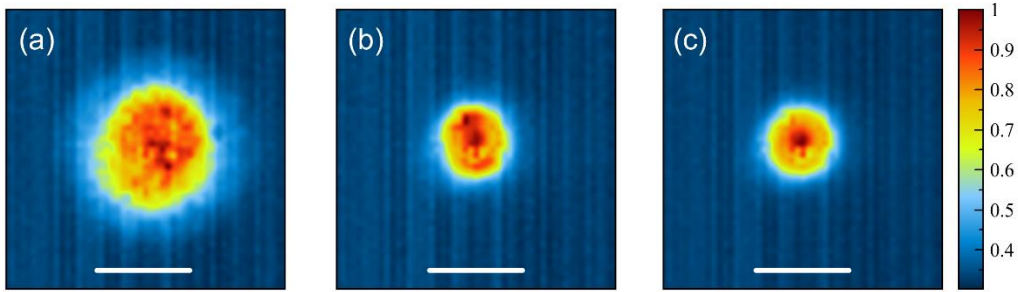

**Supplementary Figure 3.** Measured output beam profiles during spectral control experiments with 85 kW pump peak power. Scale bars indicated for measured beam profiles are 500 µm.

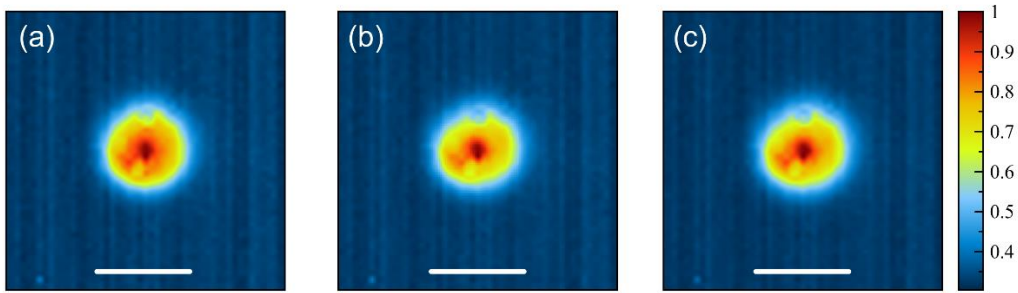

**Supplementary Figure 4.** Measured output beam profiles during spectral control experiments with 150 kW pump peak power. Scale bars indicated for measured beam profiles are 500 µm.